\documentclass[pdflatex,sn-mathphys-num]{sn-jnl}

\usepackage{graphicx}%
\usepackage{multirow}%
\usepackage{amsmath,amssymb,amsfonts}%
\usepackage{amsthm}%
\usepackage{mathrsfs}%
\usepackage[title]{appendix}%
\usepackage{xcolor}%
\usepackage{textcomp}%
\usepackage{manyfoot}%
\usepackage{booktabs}%
\usepackage{algorithm}%
\usepackage{algorithmicx}%
\usepackage{algpseudocode}%
\usepackage{listings}%

\usepackage{array}
\usepackage{threeparttable}
\usepackage{makecell,booktabs}  
\usepackage{multirow} 
\usepackage{booktabs}
\usepackage{tabulary}

\theoremstyle{thmstyleone}%
\theoremstyle{thmstyletwo}%

\theoremstyle{thmstylethree}%

\begin{document}

\title[Article Title]{A 3D-integrated BiCMOS-silicon photonics high-speed receiver realized using micro-transfer printing}


\author*[1]{\fnm{Ye} \sur{Gu}}\email{Ye.Gu@ugent.be}
\equalcont{These authors contributed equally to this work.}

\author*[2]{\fnm{He} \sur{Li}}\email{He.Li@ugent.be}
\equalcont{These authors contributed equally to this work.}

\author[1]{\fnm{Tinus} \sur{Pannier}}\email{Tinus.Pannier@ugent.be}
\equalcont{These authors contributed equally to this work.}

\author[1]{\fnm{Shengpu} \sur{Niu}}
\author[3,4]{\fnm{Patrick} \sur{Heise}}
\author[3]{\fnm{Christian} \sur{Mai}}

\author[5]{\fnm{Prasanna} \sur{Ramaswamy}}
\author[5]{\fnm{Alex} \sur{Farrel}}
\author[5]{\fnm{Alin} \sur{Fecioru}}
\author[5]{\fnm{Antonio} \sur{Jose Trindade}}
\author[5]{\fnm{Ruggero} \sur{Loi}}

\author[1]{\fnm{Nishant} \sur{Singh}}
\author[2]{\fnm{Senbiao} \sur{Qin}}
\author[2]{\fnm{Biwei} \sur{Pan}}
\author[2]{\fnm{Jing} \sur{Zhang}}
\author[6]{\fnm{Johanna} \sur{Rimböck}}
\author[7]{\fnm{Kristof} \sur{Dhaenens}}
\author[7]{\fnm{Toon} \sur{De Baere}}
\author[7]{\fnm{Geert} \sur{Van Steenberge}}
\author[8]{\fnm{Dieter} \sur{Bode}}
\author[8]{\fnm{Dimitrios} \sur{Velenis}}
\author[8]{\fnm{Guy} \sur{Lepage}}
\author[8]{\fnm{Neha} \sur{Singh}}
\author[8]{\fnm{Joris} \sur{Van Campenhout}}

\author[1]{\fnm{Xin} \sur{Yin}}
\author*[2]{\fnm{Günther} \sur{Roelkens}}\email{Gunther.Roelkens@UGent.be}
\author*[1]{\fnm{Peter} \sur{Ossieur}}\email{Peter.Ossieur@imec.be}

\affil[1]{\orgdiv{IDLab, INTEC}, \orgname{Ghent University – imec},  \orgaddress{\city{Ghent}, \postcode{9052},  \country{Belgium}}}

\affil[2]{\orgdiv{Photonics Research Group, INTEC}, \orgname{Ghent University – imec},  \orgaddress{\city{Ghent}, \postcode{9052},  \country{Belgium}}}

\affil[3]{\orgname{IHP- Leibniz-Institut für innovative Mikroelektronik},   \orgaddress{\city{Frankfurt (Oder)},  \country{Germany}}}

\affil[4]{\orgname{Technical University of Applied Sciences Mittelhessen}, \city{Gießen}, \postcode{35390}, \country{Germany}}

\affil[5]{ \orgname{X-Celeprint Ltd.}, \orgaddress{\street{Lee Maltings, Dyke Parade}, \city{Cork},  \country{Ireland}}}

\affil[6]{ \orgname{EV Group}, \orgaddress{\street{E. Thallner GmbH}, \city{St. Florian am Inn},\country{Austria}}}

\affil[7]{\orgdiv{Center for Microsystems Technology (CMST), INTEC}, \orgname{imec – Ghent University},  \orgaddress{\city{Ghent}, \postcode{9052},  \country{Belgium}}}

\affil[8]{\orgname{imec}, \orgaddress{\street{Kapeldreef 75}, \city{Leuven}, \postcode{3001}, \country{Belgium}}}


\abstract{
Meeting the escalating demands of data transmission and computing, driven by artificial intelligence (AI), requires not only faster optical transceivers but also advanced integration technologies that can seamlessly combine photonic and electronic components. Traditional approaches struggle to overcome the parasitic limitations arising from fabricating those components using different processes. Here, we report a novel 3D heterogeneously integrated optical receiver based on micro-transfer printing ($\mu$TP), enabling the co-integration of a compact bipolar CMOS (BiCMOS) electronic chiplet (0.06 \(mm^2\)) directly onto a silicon photonic integrated circuit (SiPIC). While previous $\mu$TP demonstrations have focused primarily on photonic integration, our work pioneers the direct integration of electronics and photonics, significantly enhancing performance and scalability. 
The resulting optical receiver achieves 224 Gb/s four-level pulse amplitude modulation (PAM-4) operation, delivering -5.2 dBm optical modulation amplitude (OMA) sensitivity at a bit-error rate (BER) of \(2.4\times10^{-4}\), a record-small footprint, and an excellent power efficiency of 0.51 pJ/b. This demonstration not only showcases the potential of $\mu$TP for high-density, cost-efficient integration but also represents a critical step toward next-generation optical interconnects in the AI era.

}




\maketitle


\section{Introduction}\label{sec1}



The unprecedented acceleration in  AI development is driven by remarkable advances in fields such as large language models and deep neural networks. As AI tasks continue to increase in both scale and complexity, the corresponding computational requirements have escalated dramatically. 
The implementation of optical interconnect technology enables high-bandwidth and energy-efficient communications between graphics processing units (GPUs) or central processing units (CPUs), supporting the development of scalable computing infrastructures for modern AI systems\cite{jocnNadal,ossieur2023high,jocnmaniotis,baehr2023monolithically}. Optical transceivers serve as the interface between electrical and optical signals, and determine the performance and power efficiency of optical links. Those transceivers are typically composed of photonic integrated circuits (PICs) and electronic integrated circuits (EICs) from different processes and nodes, and heterogeneous integration is needed\cite{jltjakob,jlt132G,jsscpatel,issccenrico,ofczafrany,ofcrho,neli}. Monolithic integration, which integrates photonics and electronics into a single electronic-photonic integrated circuit (EPIC) \cite{jltepic,jltpascal,jsscmovaghar}, is associated with obstacles such as fabrication complexity, material constraints, high cost and limited design flexibility.

Heterogeneous integration of PICs and EICs allows largely independent selection and optimization of the processes with which both are fabricated, leading to superior performance as long as the interconnect parasitics can be kept small. Conventional 2-dimensional (2D) integration, such as wire-bonding integration, suffers from high parasitics\cite{jorisjlt,jltye,jstqepeter,jltjakob,jsscliu}.  3D integration, leveraging flip-chip, through-silicon vias and backside redistribution layer, can reduce parasitics with shorter connections, improve integration density, and improve energy efficiency \cite{nature3D,jlt132G,jltdbi,sipchang2024}.

Micro-transfer printing ($\mu$TP) technology has emerged as a promising approach for 3D heterogeneous integration \cite{APLgunther,zhang2019iii}. Compared with the traditional flip-chip assembly, $\mu$TP offers the advantages of high throughput and low cost\cite{JSTQErevieew}. $\mu$TP allows for parallel assembly of many chiplets in a single operation, unlike flip-chip which is serial in nature. The heterogeneous integration of  PICs using $\mu$TP technology has been demonstrated; for example, the heterogeneous integration of InAs/GaAs quantum dot semiconductor optical amplifiers (SOAs) in SiPICs \cite{prliu}, lithium niobate modulators in silicon nitride PICs\cite{apltom}, and photodiodes on silicon nitride PICs\cite{aplmaes} have been successfully implemented. However, to date, no heterogeneous integration of PICs and high-speed EICs has been demonstrated to realize an optical receiver using $\mu$TP.  

The sub-micron alignment accuracy of $\mu$TP \cite{gomez2022micro} allows the EICs to occupy a small area as the X\&Y dimensions of its bondpads can be shrunk accordingly, reducing the associated electrical parasitics (capacitance) and enabling density scaling. The small footprint optical
engines are also required for co-packaged optical transceivers (CPO) with high shoreline density, so they can fit around the host application specific integrated circuits (ASICs). High-throughput integration is possible through massively parallel micro-transfer printing.  Ref. \cite{siphhe} demonstrated the $\mu$TP of an EIC on a blank silicon wafer and metallization. Here, we demonstrate a further developed $\mu$TP process to integrate a silicon–germanium (SiGe)  BiCMOS chiplet onto an advanced SiPIC, including metallization for the interconnection of both, to realize a high-speed optical receiver.

In this paper, we report the first  $\mu$TP-based 3D integrated optical receiver integrating a SiGe BiCMOS chiplet on a SiPIC. The SiGe BiCMOS chiplet and the SiPIC are connected with lithographically defined metal traces. The SiGe BiCMOS chiplet contains a transimpedance amplifier (TIA) with a small footprint of 200 \(\times\) 300 \(\mu\)\(m^2\).  224 Gb/s PAM-4 operation is demonstrated, achieving \(-5.2\) dBm OMA sensitivity at a BER of \(2.4\times10^{-4}\) (KP4-FEC) with a 6-tap feed-forward equalizer (FFE). The optical receiver achieves a low power consumption of 0.51 pJ/b and a small EIC footprint of 0.06 \( mm^2\).

\begin{figure*}[h!]
\centering
\includegraphics[width=4.5in]{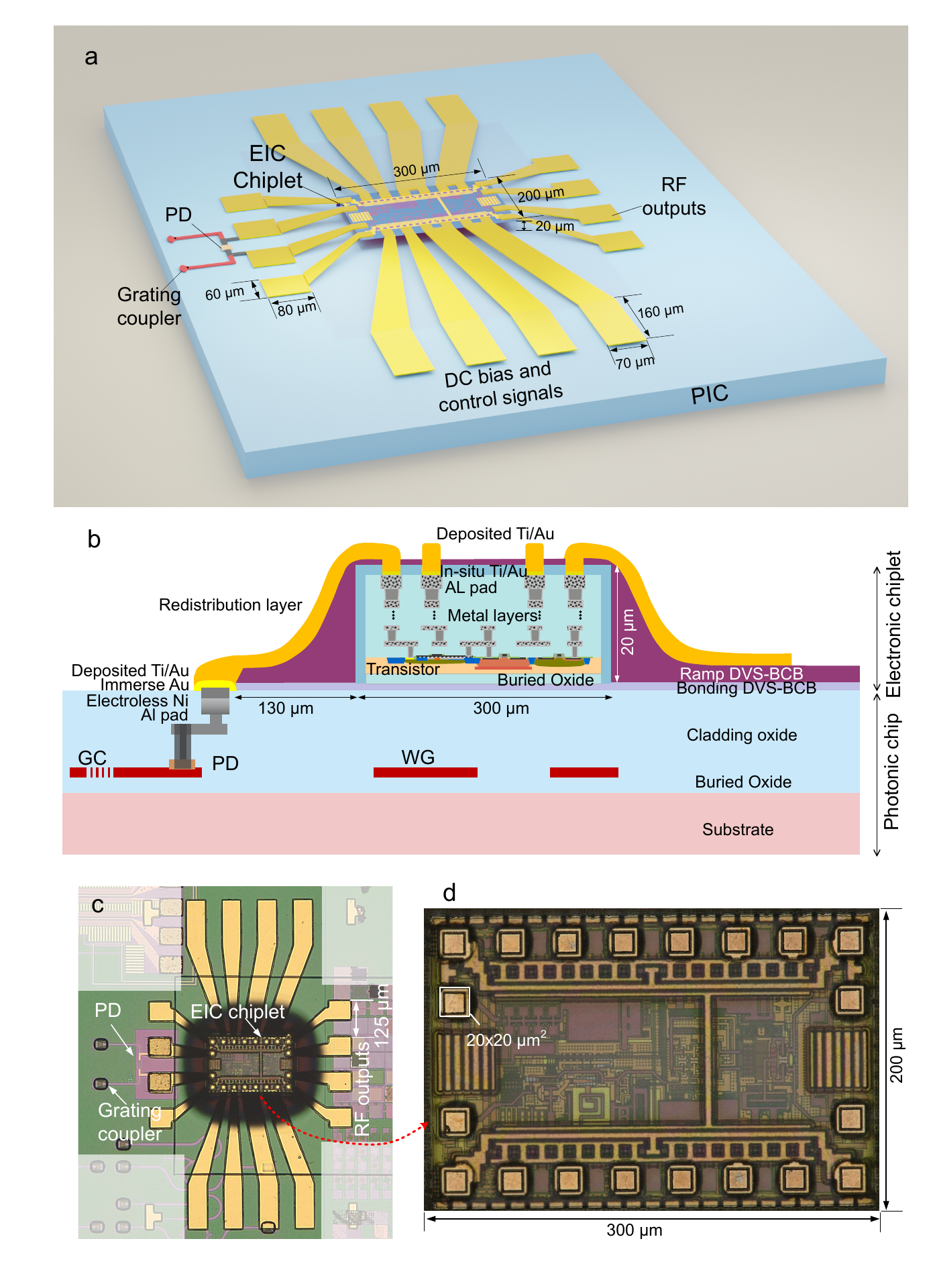}
\caption{(a) 3D diagram of the optical receiver consisting of an EIC chiplet micro-transfer printed on a SiPIC and metal traces (b) Cross-section diagram of the optical receiver (c) Micrograph of the optical receiver  (d) Micrograph of the EIC chiplet.}
\label{architecture}
\end{figure*}

\section{Concept}

The 3-D and cross-section diagrams of the $\mu$TP-based optical receiver are shown in Fig. \ref{architecture}(a) and Fig. \ref{architecture}(b), respectively. The optical receiver consists of an EIC chiplet micro-transfer printed on a  SiPIC.   Fig. \ref{architecture}(c) shows a micrograph of the realized optical receiver, and Fig. \ref{architecture}(d) shows the EIC chiplet before $\mu$TP.

The PIC consists of a grating coupler to interface with an optical fiber and a waveguide-coupled O-band Germanium photodiode (PD), which is manufactured in imec’s 300 mm silicon photonics platform (iSiPP300)\cite{ferraro2023imec}. 
Silicon photonics combines compatibility with mature CMOS fabrication technology and the ability to integrate diverse optical components, enabling the realization of compact, low-cost, and high-performance optical transceivers\cite{streshinsky}. The PD features $>$50 GHz bandwidth and 0.9 A/W responsivity at 1310 nm (O-band), supporting high baud rate and good sensitivity optical receivers. 
 

The EIC chiplet consists of a single-ended input TIA, which is used to convert the current signal from the PD into a voltage signal and to provide a buffer with 50 Ohm output impedance for measurements. The EIC is manufactured using an IHP 130$~$nm SiGe BiCMOS process.  SiGe  BiCMOS technology employs the high-speed bipolar junction transistors alongside the low-power characteristics of CMOS transistors, making it ideal for high-frequency  applications\cite{npkoch}. In this work, the SiGe BiCMOS EIC is processed on silicon-on-insulator (SOI) wafers with $<$100$>$ crystal orientation, to enable the release etch. The EIC chiplet is designed to have a compact footprint of 300 \(\mu\)m in length, 200 \(\mu\)m in width, and 20 \(\mu\)m in height, enabling high-density integration. The high alignment accuracy of  $\mu$TP technique allows the use of small pads with dimensions of 20 \(\mu\)m \(\times\) 20 \(\mu\)m, reducing the parasitic capacitance. 


The anode of the PD is connected to the input of the TIA, and the cathode of the PD is connected to a bias voltage on the TIA chiplet. As the pads of the TIA chiplet are too small to be probed or wire-bonded, the outputs of the TIA chiplet are fanned out on a divinylsiloxane-bis-benzocyclobutene (DVS-BCB) interlayer to larger pads (70 \(\mu\)m \(\times\) 80 \(\mu\)m) with a 125-\(\mu\)m pitch, which can be measured with an electrical GSSG RF probe.  The pads for the power, ground, and control signals are also redistributed to larger pads (70 \(\mu\)m \(\times\) 160 \(\mu\)m) with Au traces for electrical interconnection in the measurement. The EIC chiplet is micro-transfer printed about 130  \(\mu\)m away from the pads of the PD. It should be noted that, in order to further reduce interconnect parasitics, this distance could be shortened by further process optimizations. The pad of the PD on the PIC is 60 \(\mu\)m \(\times\) 80 \(\mu\)m, which can also be further shrunk to reduce the parasitic capacitance. The choice for 20 \(\mu\)m \(\times\) 20 \(\mu\)m and 60 \(\mu\)m \(\times\) 80 \(\mu\)m pads was made to reduce risks in this proof-of-concept demonstrator. Further development of mainly the post-print metallization process would allow for even far smaller pad sizes below 10 \(\mu\)m \(\times\) 10 \(\mu\)m on both PIC and EIC, and correspondingly smaller pitches between such pads. 

\section{Integration process}

$\mu$TP provides an efficient approach for heterogeneous integration with minimal silicon electronic chip-area usage, as the bondpad size can be drastically reduced. The overall system performance is augmented due to lower parasitics, the risks associated with monolithic EPIC chip development are mitigated, and functionality becomes increasingly modularized. This strategy results in improved performance, decreased power consumption, reduced costs, and enhanced chip yield.

\begin{figure}
    \centering
    \includegraphics[width=1\linewidth]{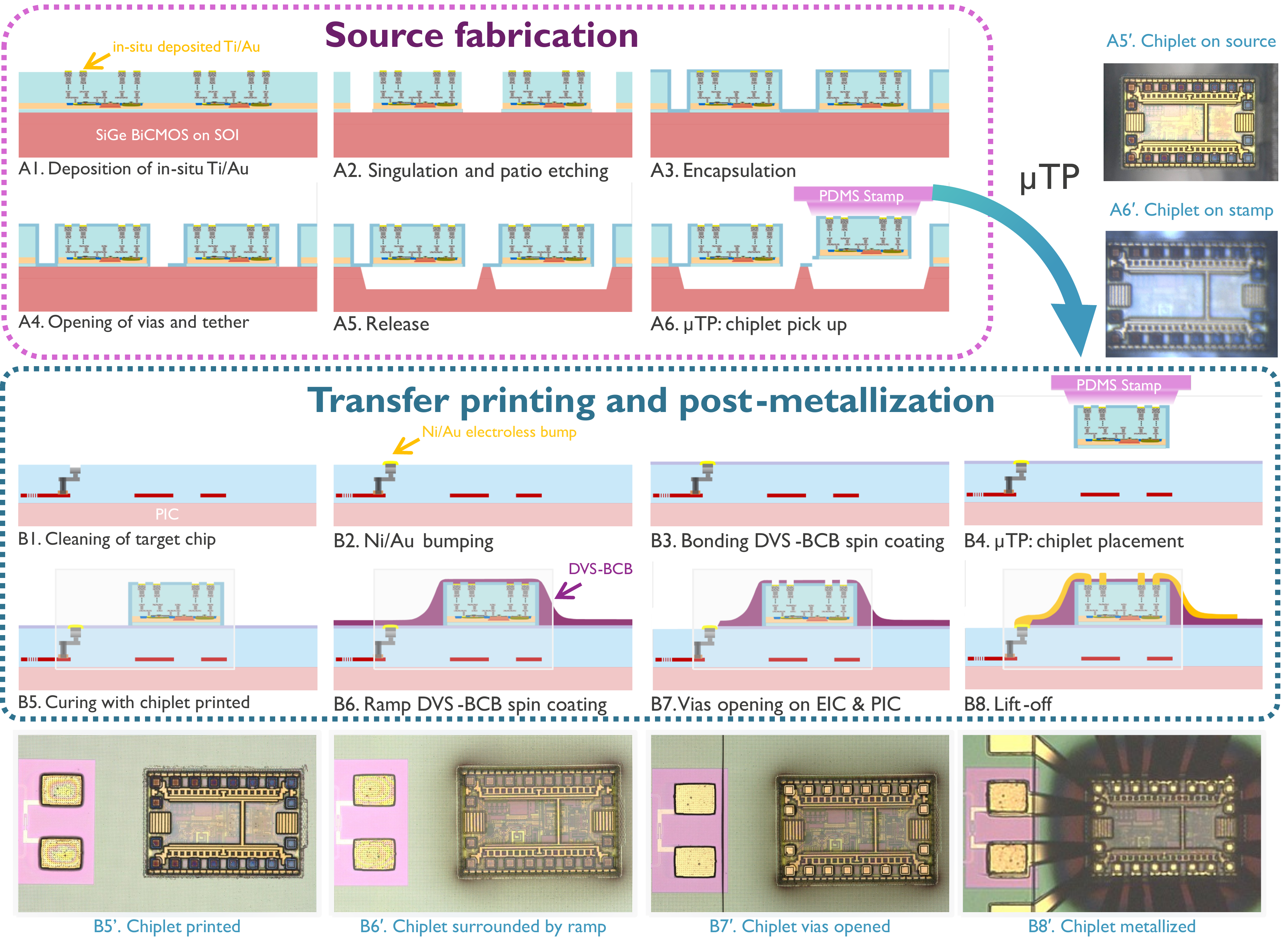}
    \caption{A1–A6, Source fabrication steps including in-situ Au deposition, chiplet singulation, encapsulation, via opening, release, and $\mu$TP pick-up. A5'-A6': top view of released chiplet on source and stamp corresponding to A5-A6.
B1–B8, Transfer printing and post-metallization steps including target cleaning, bumping, DVS-BCB coating, $\mu$TP placement, curing, ramping, via opening, and final metallization. B5'-B8': top view of chiplet printed, covered by ramp, via-opened and metallized  corresponding to B5-B8.}
    \label{fig:placeholder}
\end{figure}

The integration process is schematically depicted in Fig. \ref{fig:placeholder}, with A1–A6 corresponding to source fabrication and B1–B8 to $\mu$TP and post-metallization process. The source fabrication comprises the definition of EIC chiplets on the source wafer by etching trenches, followed by dielectric encapsulation, tether defining and undercut etching. A poly-dimethylsiloxane (PDMS) stamp is then laminated against the source wafer, followed by fast retraction of the stamp, leaving the chiplet attached to the stamp. The chiplet is aligned and laminated to the target substrate, after which the stamp is retracted. 
The target photonic substrate is deposited with the electroless Ni/Au, cleaned, and coated with a DVS-BCB adhesive bonding layer onto which the electronic chiplets are printed. After curing, the substrate is coated again with DVS-BCB to form ramps, and the DVS-BCB is etched to form vias to the contact pads. The metallization traces are then formed to connect the TIA to the PD. More details on the fabrication process can be found in the Methods section.




\section{Opto-electrical performance}

\begin{figure*}[h!]
\centering
\includegraphics[width=1\linewidth]{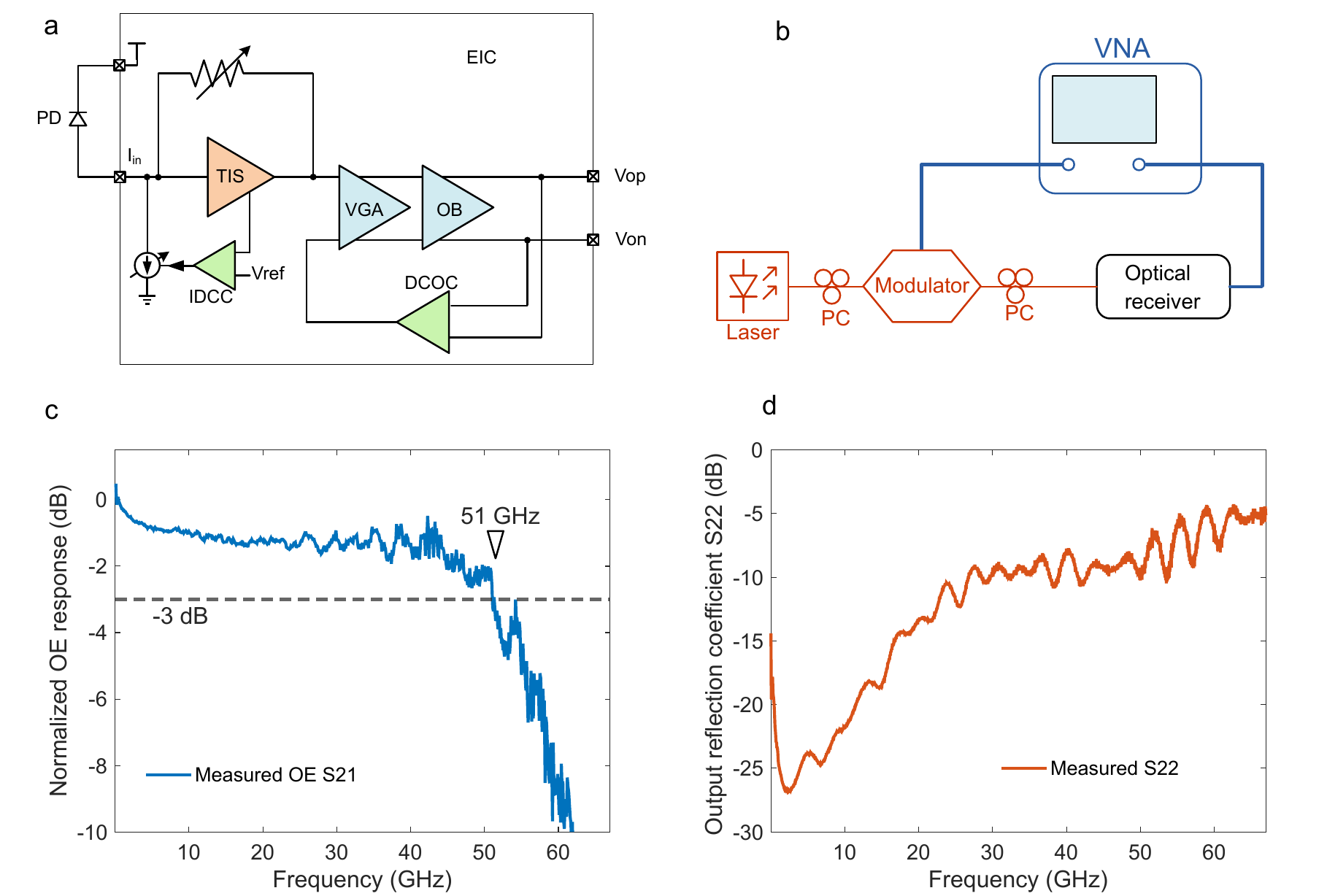}
\caption{Opto-electrical performance. (a) Block diagram of the optical receiver (b) Opto-electrical frequency response experiment setup (c) Measured OE response (d) Measured output reflection coefficient.}
\label{EIC}
\end{figure*}

To realize high-speed PAM-4 optical receivers for next-generation optical links, EICs with high bandwidth, low noise, and low power consumption are required. The block diagram of the optical receiver is shown in Fig. \ref{EIC}(a). The EIC consists of a TIA input stage (TIS), a variable gain amplifier (VGA), an output buffer (OB), input DC current cancellation (IDCC), and a DC offset cancellation (DCOC) circuit. The schematic of the EIC can be found in the Supplementary Information.

Although a traveling wave circuit can achieve a higher bandwidth compared with a lumped-element circuit, it consumes much more power and occupies a larger area\cite{jltjakob,sackinger2005broadband}. Here, we use a lumped-element circuit structure to design the high-speed TIA to realize power-efficient chiplets. A shunt feedback structure is utilized as the TIS for low noise, high bandwidth and low power consumption \cite{sackingerbook}. The IDCC circuit is used to absorb the DC current of the PD and provide a DC bias for the TIS stage. The feedback resistor is designed to be tunable to confront process variations. The digitally programmable VGA further amplifies the voltage signal of the TIS. The VGA, whose input common-mode reference voltage is generated by the DCOC loop, also serves as a single-to-differential converter. The OB with a 50 $\Omega$ output impedance is intended to send the output signals out of the chip for measurement. A tunable continuous-time linear equalizer is implemented in the OB to compensate for the loss at high frequencies. The output DC offset, caused by the VGA and OB, is eliminated by the DCOC circuit. A 2.5 V voltage is used as the power supply to realize low-power operation.

In this work, the TIA chiplet is designed with a small footprint of 300 \(\mu\)m \(\times\) 200 \(\mu\)m, allowing high-density integration. The high alignment accuracy of  $\mu$TP allows the use of small pads of 20 \(\mu\)m \(\times\) 20 \(\mu\)m (and potentially smaller), enabling a compact chiplet size. The small pads also reduce the parasitic capacitance at the input and output, allowing an increase in the bandwidth and a reduction in the power consumption.

The opto-electrical (OE) response of the optical receiver is characterized using a VNA; the experiment setup is shown in Fig. \ref{EIC}(b). The grating coupler on the PIC is probed using a fiber probe, and the outputs of the optical receiver are probed using a GSSG RF probe. The measured OE response is shown in Fig. \ref{EIC}(c); the optical receiver achieves 51 GHz 3-dB bandwidth. The measured output reflection coefficient is shown in Fig. \ref{EIC}(d). The output reflection coefficient S22 is less than -8 dB up to 50 GHz.  The OE performance demonstrates the capability of achieving high-speed operation.

\section{Data transmission}

\begin{figure*}[h]
\centering
\includegraphics[width=5in]{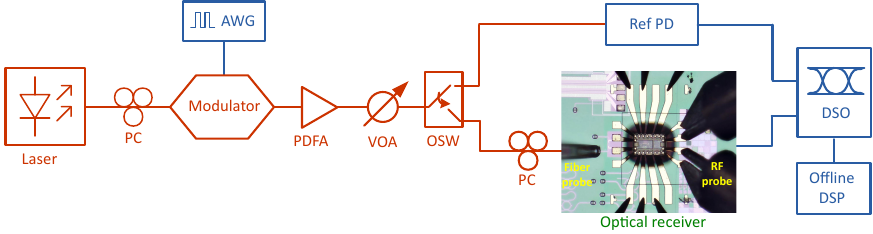}
\caption{Data transmission experiment setup }
\label{setup}
\end{figure*}

\begin{figure*}[h]
\centering
\includegraphics[width=1\linewidth]{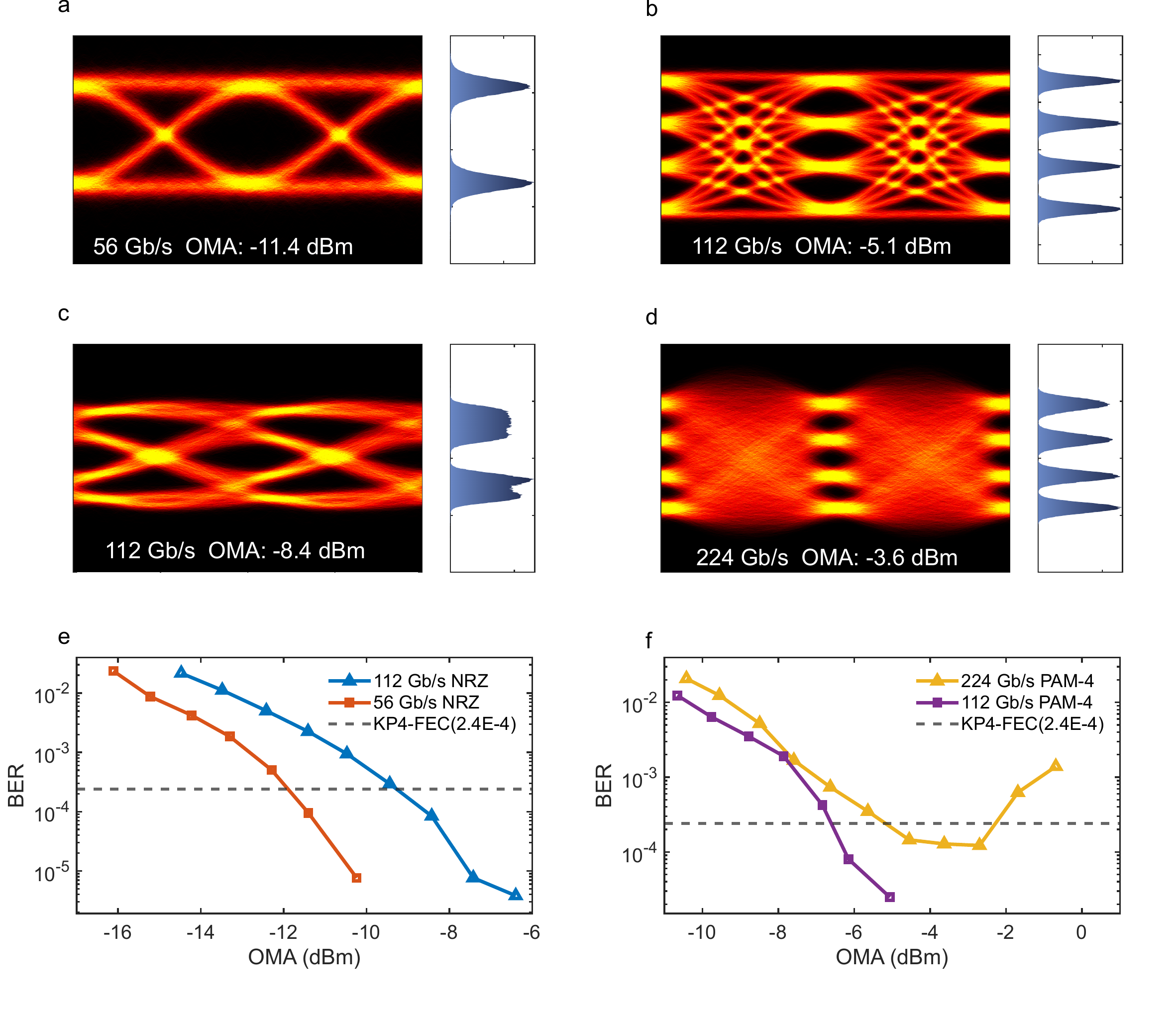}
\caption{Data transmission experiment results. (a) 56 Gb/s NRZ eye diagram measured at -11.4 dBm. (b) 112 Gb/s PAM-4 eye diagram measured at -5.1 dBm. (c) 112 Gb/s NRZ eye diagram measured at -8.4 dBm.  (d) 224 Gb/s PAM-4 eye diagram measured at -3.6 dBm with 6-tap FFE. (e) Measured 56 Gb/s and 112 Gb/s NRZ BER curves.(f) Measured 112 Gb/s and 224 Gb/s PAM-4 BER curves.}
\label{eyediagram}
\end{figure*}

A data transmission experiment is carried out to characterize the time-domain performance of the optical receiver. The measurement setup is shown in Fig. \ref{setup}. On the transmitter side, the laser is modulated by a \(\ 2^{13}-1\) pseudo-random bit sequence (PRBS) pattern generated by an arbitrary waveform generator (AWG). A 70 GHz reference PD (Finisar XPDV3120) is used to capture the eye diagrams and the extinction ratios of the transmitter. BERs are obtained using the error bit counting method in MATLAB based on the recorded waveforms. 

The 56 and 112 Gb/s non-retern-to-zero (NRZ) eye diagrams of the optical receiver measured by the digital sampling oscilloscope (DSO) are shown in Fig. \ref{eyediagram} (a) and (c). All the NRZ eye diagrams are open without any equalization on the scope. The measured BER curve is also shown in Fig. \ref{eyediagram} (e). Each BER point in the BER curves has at least 10 error bits in the BER counting. The optical receiver achieves -11.9 and -9.2 dBm optical modulation amplitude (OMA) sensitivity at the KP4-forward-error-correction (FEC) threshold of \(2.4\times10^{-4}\) for 56 and 112 Gb/s NRZ modulations, respectively. It should be noted that the minimum measured BER is limited by the finite waveforms recorded by the DSO.

The 112 and 224 Gb/s PAM-4 eye diagrams of the optical receiver measured by the DSO are shown in Fig. \ref{eyediagram}(b) and (d). 6-tap FFEs are employed on the scope to open the 224 Gb/s PAM-4 eye diagrams of the optical receiver. The 112 Gb/s PAM-4 eye diagram is widely open without any equalization used in the scope. The measured PAM-4 BER curves are also shown in Fig. \ref{eyediagram}(f). The optical receiver achieves -6.6 and -5.2 dBm OMA sensitivity at the KP4-FEC threshold of \(2.4\times10^{-4}\) for 112 and 224 Gb/s PAM-4 modulations, respectively. The total power consumption is 115 mW, corresponding to an energy efficiency of 0.51 pJ/b for 224 Gb/s PAM-4 operation.

\section{Discussion}

The measurement results demonstrate that the $\mu$TP-based optical receiver exhibits a high bandwidth, achieving 224 Gb/s PAM-4 operation. To the best of our knowledge, this is the first demonstration of a $\mu$TP-based optical receiver. The compact footprint of the EIC chiplet provides the potential for the integration of several chiplets on a single PIC to realize multi-channel IMDD receivers to increase the total data rate. This technique can also be used to realize other optical receivers, for example, coherent receivers and self-coherent receivers, which can utilize advanced modulation formats to achieve higher bit rates per lane.

Table I in the supplementary information shows the comparison with state-of-the-art IMDD optical receivers  using different integration methods. Ref. \cite{jltjakob} reported a 224 Gb/s optical receiver using 2D wirebonding integration, where a traveling-wave transimpedance amplifier has been used to compensate the high parasitic inductance of the bond wires, which results in a relatively high power consumption of 2.2 pJ/b. Ref \cite{jlt132G} reported a 264 Gb/s optical receiver using 3D direct-bonding integration, but it still features a power consumption of 1.45 pJ/b and occupies a relatively large EIC chip area of 0.6 \(mm^2\). This work achieves the lowest power consumption of 0.51 pJ/b and the smallest EIC area of 0.06 \(mm^2\) among \(>\)200 Gb/s PAM-4 optical receivers.  Our work provides a new integration method for next-generation high-speed optical receivers. Heterogeneous integration realized by $\mu$TP signifies a substantial advancement towards realizing multiterabit-per-second optical interconnects. Such systems enable compact and energy-efficient designs for next-generation computing and communication applications by integrating photonic and electronic functionalities into a unified platform. Such integration strategies establish a foundation for scalable and cost-effective solutions that meet the high-performance demands of modern systems and foster innovation across various application domains.

\section{Methods}

\subsection{Sample preparation}

The EICs are manufactured at IHP (Germany) using a tailored version of the SG13G2 SiGe BiCMOS process \cite{rucker2012half}. Rather than relying on the bulk silicon wafers of the SG13G2 technology, the process flow is modified to be compatible with a 200 mm SOI handle wafer with $<$100$>$ crystal orientation incorporating a 1-$\mu$m buried oxide (BOX) layer and a 1.5-$\mu$m top silicon device layer for the front end of line (FEOL). The back end of line (BEOL) has a height of about 15 $\mu$m and consists of seven metal layers, i.e. five thin Al-layers and two thick Al-layers. The chiplets are arranged on a dense grid on the reticle. To facilitate the release process, during the mask design it is ensured no metals are present in the spaces between the chiplets.

The processing of the EIC source chip to enable the micro-transfer printing is explained in \cite{loi2022micro}. After completing the wafer in IHP, Ti/Au is deposited to protect the contact pads and also serves as the process monitor in the following etching processes in X-Celeprint. Then, the chiplets are singulated by inductively coupled plasma (ICP) with a C\textsubscript{4}F\textsubscript{8}/H\textsubscript{2}/He dry-etching to etch through the BEOL oxide stack and a SF\textsubscript{6}/C\textsubscript{4}F\textsubscript{8} dry-etching to etch through the FEOL silicon. With another lithography to define the encapsulation patio, the remaining buried oxide (BOX) is etched away by C\textsubscript{4}F\textsubscript{8}/H\textsubscript{2}/He, isolating all the coupons. Then the substrate is covered by a dielectric double layer of 500-1000 nm SiO\textsubscript{2} and 1.5-2~\(\mu\)m SiN. Another lithography and C\textsubscript{4}F\textsubscript{8}/H\textsubscript{2}/He ICP dry-etch are used to define the tethers and the opening of the contact pads on the chiplets. Undercut and release of the coupons are achieved using a tetramethylammonium hydroxide (TMAH) solution at a ratio of TMAH:DI (1:4).\

The target PIC is manufactured using imec's iSIPP300 process, a 300 mm photonic platform for high-speed applications. In the first post-processing step, The target PIC is cleaned using standard wet cleaning recipe with acetone, isopropyl alcohol, and deionized water, and then the aluminum contact pads are electroless-plated with Ni/Au to avoid the oxidation of aluminum and ensure low-ohmic contact during post-printing metallization. The sample is then cleaned again with wet cleaning. After oxygen plasma, the sample is spin coated with AP3000 as the adhesion promoter for DVS-BCB.  DVS-BCB is spin coated and soft cured at 150~$^\text{o}$C for 15 min. The EIC chiplet is then transfer printed using a PDMS stamp and placed close to the PD with high placement accuracy of 1 \(\mu\)m 3-sigma with micro-TP100 transfer printer. The DVS-BCB was then fully cured in a vacuum oven.
Next, 2-$\mu$m DVS-BCB is spin-coated to form a ramp for metallization followed by a complete cure. Subsequently, the chip is spin-coated with a bilayer of AZ10XT resist to achieve thick coverage over a large topography of more than 20 $\mu$m. 
The resist over the contact pads on both the EIC and PIC is exposed and developed. Reflow of the resist is used to create a sloped sidewall on the via to make metallization easier.
Dry etching is used to etch the DVS-BCB and the remaining dielectrics. Next, a Ti35E bilayer is used for better coverage for the lift-off lithography.
The metal traces are then formed through the evaporation of Ti/Au and defined with a conventional lift-off process.

\subsection{Opto-electrical experiment }

A laser (Santec TSL-570) generates a 1310 nm 13 dBm optical signal, which is coupled to an O-band lithium niobate Mach–Zehnder Modulator (MZM). The MZM is modulated by one port of a vector network analyzer (VNA) (Keysight PNA-X N5247B). The modulated optical signal is coupled to the optical receiver through an optical fiber. The differential outputs of the optical receiver are probed using a GSSG electrical probe with a 125 $\mu$m pitch and connected to the ports of the VNA. After the measurement, the S-parameters of the modulator and the electrical probe are de-embedded to obtain the normalized OE response of the optical receiver.

\subsection{Data transmission experiment}

A 1310 nm 13 dBm optical signal is generated by a continuous-wave (CW) laser (Santec TSL-570). A 256 GSa/s AWG (Keysight M8199B) generates a \(\ 2^{13}-1\) PRBS signal and drives an O-band lithium niobate MZM directly. The MZM is biased at the quadrature point by a DC supply, followed by a praseodymium-doped fiber amplifier (PDFA) to increase the optical power. A variable optical attenuator (VOA) is used to control the optical power coupled to the PD. An optical switch (OSW) is followed by a reference PD with a 70 GHz bandwidth and a polarization controller.  The output of the reference PD is connected to the remote sampling head of a DSO (Keysight N1046A).  To compensate for the bandwidth limitation at the transmitter side, de-embedding is utilized in the AWG to obtain a clear eye diagram at the output of the transmitter measured through the reference PD. The grating coupler on the PIC is probed by a fiber probe, and the coupling efficiency is improved by a polarization controller (PC). The outputs of the TIA chiplet are probed by a 67 GHz GSSG electrical probe with a pitch of 125 $\mu$m and measured by two remote sampling heads of the DSO. The pads of power, ground, and control signals on the PIC are connected with DC probes. BERs are obtained using offline data processing in MATLAB based on the recorded waveforms. The OMAs are calculated based on the average currents of the PD on the PIC measured by a current meter and the extinction ratios after the MZM measured from the reference PD.

\backmatter

\bmhead{Supplementary information}

Supplementary information is available.

\bmhead{Acknowledgements}

This work is supported by EU-funded project \mbox{CALADAN} under grant number 825453. 

\bmhead{Author contribution}

Y.G. designed the electrical IC and measured the receiver. H.L. developed the micro-transfer printing and metallization process. T.P. designed the photonic IC and helped with the measurement. S.N. helped with electrical IC design. P.H. and C.M. helped with the electrical IC fabrication. P.R., R.L., A.F., A.T. and A.F. performed the release of chiplets. N.S. helped with the measurements. S.Q., B.P., and J.Z. helped with the micro-transfer printing. K.D., T.D., and G.V. assisted with electroless plating of nickel and gold. J.R. performed the spray-coating. D.B., D.V., G.L., N.S., and J.C. helped with the photonic IC fabrication. X.Y., G.R. and P.O. supervised the work. Y.G. and H.L. prepared the manuscript.


\bibliography{sn-bibliography.bib}

\newpage

\section{Supplymentary}

\subsection{Transimpedance amplifier chiplet}\label{sec1}

\begin{figure*}[h!]
\centering
\includegraphics[width=5.2in]{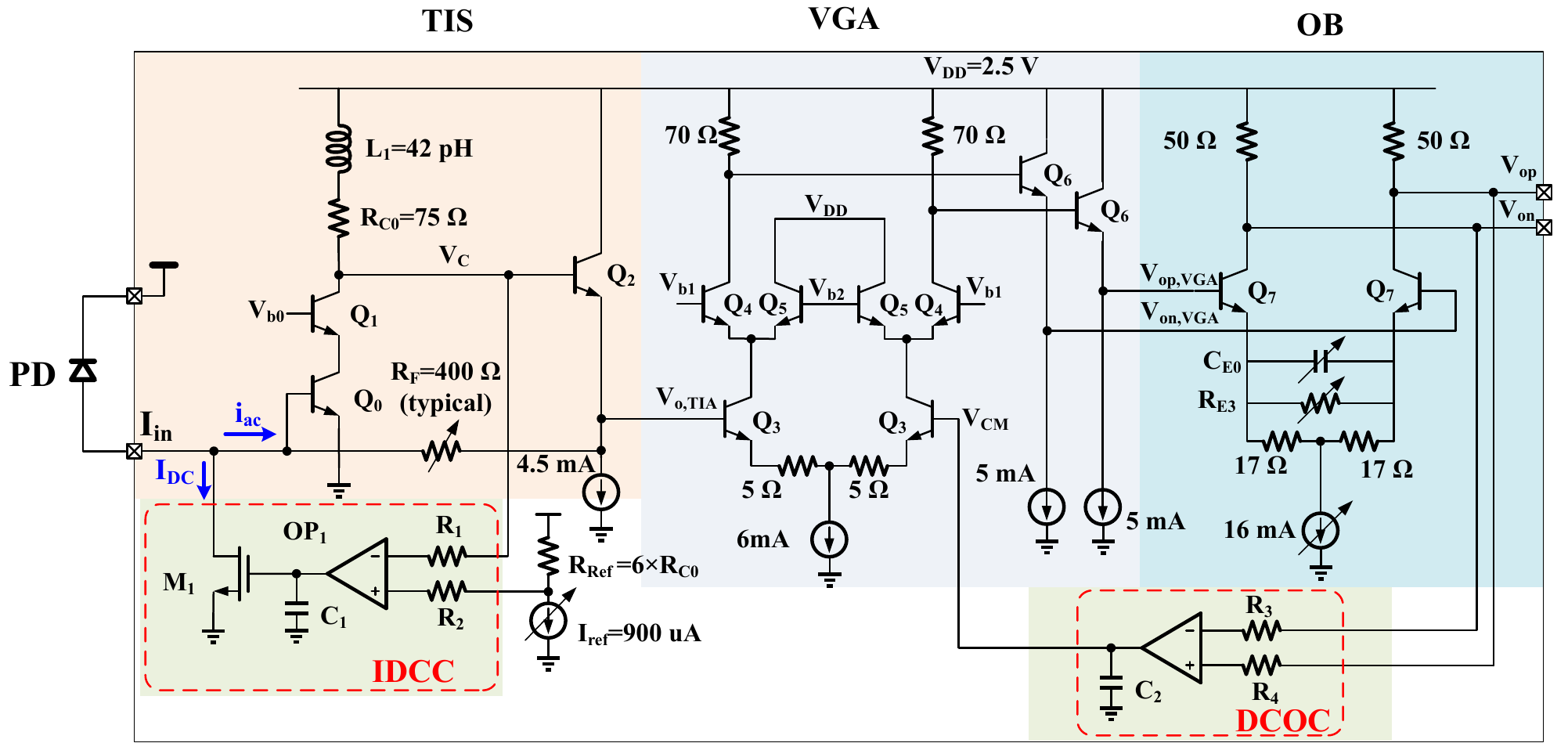}
\caption{Simplified schematic of the TIA chiplet.}
\label{schematicTIA}
\end{figure*}

The simplified schematic of the transimpedance amplifier (TIA) chiplet is shown in Fig. \ref{schematicTIA}. The TIA  input stage (TIS) utilizes a typical shunt feedback structure to convert the AC current signal of the PD into a voltage signal. Two electrostatic discharge (ESD) diodes, with a total of 51 fF parasitic capacitance, are placed at the input of the TIS to protect the internal circuits. $Q_1$ serves to alleviate the Miller effect of the parasitic base-collector capacitor of $Q_0$. A shunt inductor $L_1$, with 25\(\times\)25 \(\mu\)m\(^2\) area, is used to increase the phase margin of the TIA loop. The feedback resistor $R_F$, implemented as NMOS transistors and a polysilicon resistor,  can be adapted by digital bits to compensate for the process variation. \(M_1\) in the input DC current cancellation (IDCC) is used to absorb the DC current of the PD and the DC current of $R_F$. \(C_1\) is used to compensate for the IDCC loop and alleviate the baseline wander. $R1=4 k\Omega$, placed close to the base of $Q_2$, is used to prevent the parasitic capacitance of the IDCC from affecting the high-speed path. As \(R_{Ref}=6\times R_{C0}\), the $Q_0$  collector current $I_{C,Q_0}\approx 6 \times I_{ref} = 5.4  mA$,  which impacts the overall bandwidth and input-referred noise current of the TIA. $I_{ref}$ is designed to be tunable and controlled by digital registers. $R_{Ref}$ and $R_{C0}$ are placed close to each other in the layout for matching.

The DC voltage at the input of the TIS is equal to one base-emitter forward bias voltage $V_{be,Q_0}\approx0.75 V$ .  The PD cathode bias voltage can be chosen to be the same as the main supply voltage of 2.5 V, so that the PD is reversely biased with a $\sim$1.75 V voltage difference.

The TIS is followed by the variable gain amplifier (VGA) stage, which also serves as a single-to-differential converter. One input of the VGA is connected to the output of the TIS, the other input is connected to a reference voltage generated by the DCOC. The gain of the VGA can be fine-tuned by changing $V_{b1}$ and $V_{b2}$ through digital bits.

 Two 50 $\Omega$ resistors are used for impedance matching in the output buffer (OB) stage. The OB stage has a tunable CTLE, realized by a tunable capacitor $C_{E0}$ and a tunable resistor $R_{E3}$. The $R_{E3}$ is realized using NMOS transistors in the triode region, whose resistance is programmable by tuning their gate voltages. The $C_{E0}$ is realized using PMOS varicaps, which can be fine-tuned by digital registers. The differential outputs of the OB are also protected by ESD diodes.

\newpage

\subsection{Comparison with current state-of-the-art optical receivers using different integrations}

Table \ref{tab1} shows the comparison with the state-of-the-art IMDD optical receivers  using different integration methods. This $\mu$TP based optical receiver achieves the lowest power consumption of 0.51 pJ/b and the smallest EIC area of 0.06 \(mm^2\).


\begin{table*}[h!]
    \centering

    \caption{Comparison with current state-of-the-art PAM-4 optical receivers using different integrations.}
    \label{tab1}
    \footnotesize	
    \renewcommand{\arraystretch}{0.5}
    \setlength{\tabcolsep}{4pt}  

    \begin{threeparttable}[b]

    \begin{tabular}{|c|c| c|c|c|c|c|c|c|c|c}
         \hline
        ref. & \makecell{Integration} &  \makecell{EIC\\technology} & \makecell{Data\\rate\\(Gb/s)} & \makecell{\(R_T\)\\(dB\(\Omega\))} 
         & \makecell{R\\(A/W)} & \makecell{Sens.\\@KP4\\(dBm)} & { pJ/b } &\makecell{EIC\\ Area\\ (\( mm^2\))}  & DSP \\
         \hline
         
        \makecell{\cite{jltye}}& \makecell{Wire-bonding\\(2D)} & \makecell{130nm\\ BiCMOS}   & 160  & 65  & 0.8 & -7 & 0.99& 0.82\tnote{1}  & 5-tap FFE  \\
         \hline

        \makecell{\cite{jltjakob} } &  \makecell{Wire-bonding\\(2D)} & \makecell{55nm\\ BiCMOS}   & 224  & 57.3 & 0.89 & -4.8 & 2.2 &1.38\tnote{1}   & 10-tap FFE  \\
         \hline
        \makecell{\cite{wuoe}} & \makecell{Flip-chip\\(3D)} & N/A & 160  & N/A & 0.85 & -2.7 & 1.2 & N/A   &51-tap FFE  \\
         \hline
        \makecell{\cite{lakisscc}} & \makecell{Flip-chip\\(3D)} & \makecell{16nm\\ FinFET}  & 106.25  & 77 & N/A  & -13.97 & 0.98 & 0.64  &\makecell{12-tap FFE +\\ 1-tap DFE }  \\
         \hline

        \makecell{\cite{jlt132G}} & \makecell{Direct-bonding\\(3D)} & \makecell{130nm\\ BiCMOS}  & 264  & 60 & 0.9 & N/A & 1.45 & 0.6 & 6-tap FFE\\
         \hline

        \makecell{\cite{OEthomas}} & \makecell{Monolithic} & \makecell{45nm\\CMOS}  & 112  & 68 & 0.245 & N/A & 1 & N/A & N/A\\
         \hline

        {\makecell{This\\work}} &{\makecell{\(uTP\)\\ (3D)}}&{\makecell{130nm\\ BiCMOS}} & 224 & 53.6& 0.9 & -5.2&0.51 & 0.06   & 6-tap FFE \\

         \hline

    \end{tabular}
    		\begin{tablenotes}
				\footnotesize
				\item[1] \textit{data path core of one channel.}

			\end{tablenotes}

    \end{threeparttable}

\end{table*}


\end{document}